\documentclass[runningheads]{llncs}

\usepackage{graphicx}
\usepackage{cite}
\usepackage{amssymb,amsfonts}
\usepackage{xcolor}
\usepackage{multirow}
\usepackage[caption=false]{subfig}
\usepackage[algo2e]{algorithm2e} 
\usepackage[hyphens]{url}
\usepackage{hyperref}

\begin{document}

\title{MSPP: A Highly Efficient and Scalable Algorithm for \textbf{M}ining \textbf{S}imilar \textbf{P}airs of \textbf{P}oints}

\titlerunning{MSPP: Mining Similar Pair of Points}

\author{Subrata Saha \inst{1} \orcidID{0000-0003-2991-0481} \and 
Ahmed Soliman\inst{2}\orcidID{0000-0002-4339-1385} \and 
Sanguthevar Rajasekaran\inst{2}\orcidID{0000-0002-0137-4843}}

\authorrunning{S. Saha et al.}

\institute{Healthcare and Life Sciences Division, IBM Research, Yorktown Heights, NY 10598, USA\\ \email{subrata.saha@uconn.edu} \and University of Connecticut, Department of Computer Science and Engineering, Storrs, CT 06269, USA\\ \email{\{ahmed.soliman,sanguthevar.rajasekaran\footnote{Corresponding author}\}@uconn.edu}}

\maketitle              
\begin{abstract}
The closest pair of points problem or closest pair problem (CPP) is an important problem in computational geometry where we have to find a pair of points from a set of points in metric space with the smallest distance between them. This problem arises in a number of applications, such as but not limited to clustering, graph partitioning, image processing, patterns identification, and intrusion detection. For example, in air-traffic control, we must monitor aircrafts that come too close together, since this may potentially indicate a possible collision. Numerous algorithms have been presented for solving the CPP. 
The algorithms that are employed in practice have a worst case quadratic run time complexity. In this article we present an elegant approximation algorithm for the CPP called ``MSPP: \textbf{M}ining \textbf{S}imilar \textbf{P}airs of \textbf{P}oints." It is faster than currently best known algorithms while maintaining a very good accuracy. The proposed algorithm also detects a set of closely similar pairs of points in Euclidean and Pearson’s metric spaces and can be adapted in numerous real world applications, such as clustering, dimension reduction, constructing and analyzing gene/transcript co-expression network, among others.


\end{abstract}
\section{Introduction}

Given a set of $n$ points in any metric space, the problem of finding the closest pair of points is known as the Closest Pair Problem (CPP) and has been well studied. Rabin \cite{RAB76} proposed a randomized algorithm with an expected run time of $\mathcal{O}(n)$ where the expectation is in the space of all possible outcomes of coin flips made in the algorithm. Rabin's algorithm used the floor function as a basic operation. In 1979, Fortune and Hopcroft \cite{FH79} presented a deterministic algorithm with a run time of $\mathcal{O}(n\log{}\log{} n)$ assuming that the floor operation takes $\mathcal{O}(1)$ time. Both of these algorithms assume a $\mathcal{O}(1)$ dimensional space. The run times of these algorithms have an exponential dependency on the dimension. Other classical algorithms include \cite{PS86,SY95}. Yao \cite{YO91} has proven a lower bound of $\Omega(n\log{} n)$ on the algebraic decision tree model for a space of any dimension. This lower bound holds under the assumption that the floor function is not allowed. 

Time Series Motif Mining (TSMM) is a crucial problem that can be thought of as CPP in a large dimensional space. In one version of the TSMM problem, we are given a sequence $S$ of real numbers and an integer $\ell$. The goal is to identify two subsequences of $S$ of length $\ell$ each that are the most similar to each other (from among all pairs of subsequences of length $\ell$ each). These most similar subsequences are referred to as {\em time series motifs}. Let $C$ be a collection of all the $\ell$-mers of $S$. An $\ell$-mer is nothing but a contiguous subsequence of $S$ of length $\ell$. Clearly, the $\ell$-mers in $C$ can be thought of as points in $\Re^\ell$. As a result, the TSMM problem is the same as CPP in $\Re^\ell$. Any of the above mentioned algorithms can thus be used to solve the TSMM problem. A typical value for $\ell$ of practical interest is several hundreds (or more). For these values of $\ell$, the above algorithms (e.g. \cite{RAB76,FH79,PS86,SY95}) will take an unacceptable amount of time (because of the exponential dependence on the dimension). Designing an efficient practical and exact algorithm for the TSMM problem remains an ongoing challenge.

Mueen \emph{et al.} have presented an elegant exact algorithm called Mueen-Keogh (MK) for TSMM \cite{AEQSB09}. MK improves the performance of the brute-force algorithm with a novel application of the triangular inequality. Zhu \emph{et al.} have introduced the STOMP algorithm\cite{ref_stomp} that exploits overlaps of sequences in a time series and reuses computations from previous sequences to improve the performance of the distance matrix computation. Although in practice STOMP achieves improvement in performance, the time complexity is still quadratic i.e., $\mathcal{O}(n^{2})$ where $n$ is the length of the time series. A number of probabilistic as well as approximate algorithms are also known for solving this problem (e.g. \cite{PMPS08,BES03,TCM07,JJMY08,DCIT07,SG04,YKK05}). For instance, the algorithm of \cite{BES03} exploits algorithms proposed for finding $(\ell,d)$-motifs from biological data. The idea here is to partition the time series data into frames of certain width. Followed by this, the mean value in each frame is computed. This mean is quantized into four intervals and as a result, the original time series data is converted into a string of characters from an alphabet of size 4. Finally, any $(\ell,d)$-motif finding algorithm is applied on the transformed string to identify the time series motifs.
 
Cai \emph{et al.} \cite{ref_jump} proposed a deterministic algorithm to solve the TSMM problem, called JUMP, which outperforms existing $\mathcal{O}\left(n^2\right)$ methods by a factor of up to 100. This was done by skipping unnecessary comparisons and multiplication operations. Although this algorithm performs well in practice, especially as the length of the timeseries $n$ increases, the speed-up is dependent on the skipping fraction. The skipping fraction is defined to be the unnecessary operations which can be skipped. Li and Lin \cite{ref_ll} have recently presented a novel algorithm called LL to solve the TSMM problem in an expected $\mathcal{O}\left(n\right)$ time. The idea is to build and maintain a data structure called {\em grids}. Although the LL algorithm runs in an expected linear time, in practice the performance significantly drops if the input time series has a large number of very close pairs of points. This is because similar points end up in the same or neighboring grids and thus, requiring a lot of pair-wise distance computations.

In this article, we propose an efficient and scalable algorithm to detect a set of highly similar pairs of points in both of Euclidean and Pearson's metric spaces. It identifies the closest pair of points with a very high probability. MSPP is an out-of-core algorithm and consequently, can work on a very large high dimensional dataset with a small memory footprint. It needs two passes over the entire data residing in the disk. In the first pass, it discretizes the attribute values, encodes them with a small number of bits, and detects correlated pairs of points in the transformed space. It then computes the actual distances between pair of points identified in the previous pass by iterating over the entire dataset. The expected run time of MSPP is  $\mathcal{O}\left( n^{1+\epsilon}\log{} n\right)$ where $0 < \epsilon < 1$. In practice it is very fast. We have done rigorous experiments using both real and synthetic datasets to demonstrate its effectiveness in finding highly correlated pairs of points and its suitability in numerous real-world applications. 

The rest of this paper is organized as follows: Section~\ref{method} describes the proposed algorithm MSPP. Analyses of time complexity and accuracy of MSPP are also presented in this section. Section~\ref{results} describes the datasets used, performance metrics, and simulation results with discussions. Two real-world applications of MSPP besides closest pair detection are discussed in Section~\ref{apps}. Section~\ref{conclusion} concludes the paper. 

\section{Methods}

\label{method}

MSPP is an out-of-core algorithm. So, it can work with a very large dataset with a very small memory footprint. To start with the input is in the disk. There are two basic steps in the algorithm and in each step, we do one pass through the input data and thus, there are a total of two passes through the data. In each pass the algorithm incrementally retrieves information embedded in the dataset and after the final pass, it outputs a set of pairs of similar points. We claim that a pair of points in this set is the closest pair with a very high probability. Our algorithm works for a variety of metrics including the Euclidean distance and Pearson’s correlation.  Next, we illustrate our proposed algorithm MSPP.

\subsection{Our Algorithm MSPP}

Our algorithm works by discretizing the continuous attribute values of the points. If two points are very similar in the Euclidean space, then they also can be expected to be very similar when we transform the points using a subset of the attributes. In the first pass, we discretize the attribute values and detect the highly similar pair of points in the transformed space. In the second pass, we compute the Euclidean distance (or, Pearson's coefficient) between every pair of points found in the first pass and output a set of similar pairs of points. The details of our algorithm are provided next.

\subsubsection{First pass.} Let the input points be $p_{i}$ where $1 \leq i \leq m$. Each input point has $n$ attributes and let the attributes be  $a_{j},$ $1 \leq j \leq n$. We assume that the points are input in column-major order in the disk. Specifically, the input points are stored as a $n\times m$ matrix $M$ where each column $i$ of $M$ corresponds to a point and each row $j$ of $M$ corresponds to an attribute. In the first pass, we retrieve each row $j$ of $M$ at a time. There are two basic steps involved here:

\emph{\textbf{Discretizing and encoding attribute values:}} As stated above, row $j$ of $M$  has the values of the attribute $a_{j}$ for all the input points. There are $n$ iterations, one for each row of $M$. In iteration $j$ we retrieve row $j$ (i.e. a line in the file) from the dataset residing in the disk, $1\leq j\leq n$. Let $v_{i}^{j}$ be the value of the attirbute $a_j$ for the point $p_i$, $1\leq j\leq n,~1\leq i\leq m$.  In iteration $j$ we cluster the set of values $v_{i}^{j}$ of the attribute $a_{j}$ into $k$ disjoint clusters ($k$ being user defined). We employ $\verb!k-means++!$ clustering algorithm to perform this task because of its simplicity and expected linear time complexity. It initializes the cluster centers before proceeding with the standard $k$-means optimization iterations. With the $\verb!k-means++!$ initialization, the algorithm is guaranteed to find a solution that is $\mathcal{O}(\log{}k)$ competitive to the optimal $k$-means solution. After clustering, each value $v_{i}^{j}$ of the attribute $a_{j}$ falls into one of the $k$ clusters. We encode each $v_{i}^{j}$ with $k$ binary bits. Only one bit will be turned ``on" out of the $k$ bits. We can think of a bit as a binary variable, having two possible values called ``true'' and ``false'' where ``on'' bit contains ``true" value and the rest of the $k-1$ bits contain ``false" values. In this scenario, the ``on'' bit corresponds to that cluster where the particular attribute value $v_{i}^{j}$ of a point $p_{i}$ falls into. 

Consider the following illustrative example. Assume that the number of clusters is 3 and so, each attribute value will be encoded  with 3 bits. If an attribute value falls into the $2^{nd}$ cluster, we will turn ``on" the $2^{nd}$ bit and the rest will be turned ``off." After passing the entire dataset residing in the disk we will have encoded the $v_{i}^{j}$ values of all the attributes $a_{j}$ across all the points $p_{i}$. At the end of this pass, each point $p_{i}$ will have $n \times k$ coordinates in the binary space and the entire dataset will be represented by $m \times n \times k$ bits. Let the word length of the computing machine used be $d$ bits (usually $d=64$ in modern computing machines). If we assume that each attribute value occupies one word, then, there will be a $\frac{d}{k}$ factor reduction of the entire dataset after encoding!

\emph{\textbf{Mining similar pairs of points in the binary space:}} Next, we randomly sample a subset of the coordinates in the encoded binary space and hash the points based on this subset. Two points will be hashed into the same bucket if they have the same values for the randomly chosen coordinates. If two points fall into the same bucket in the hash table, this is a candidate pair. We keep a priority queue $Q$ that stores the best $r$ pairs that have been encountered thus far ($r$ being user defined). The key used for any pair in $Q$ will be the Hamming distance between them (across all the coordinates). The Hamming distance between each candidate pair will be computed and inserted into $Q$ if this Hamming distance is less than the largest key in $Q$. We repeat this process of sampling and hashing $t$ times (for some suitable value of $t$). In each stage of sampling the candidate pairs generated are used to update $Q$. A similar pair of points in terms of Hamming distance may not necessarily be similar in the original Euclidean space. This is the reason why we keep a priority queue $Q$.

\subsubsection{Second pass.} In the second pass we compute the Euclidean distance or Pearson's correlation coefficient between every pair of points found in $Q$ and output a set of $s$ ($s$ being user defined) best pairs. Please, note that the original dataset always resides in the disk. A pseudocode for MSPP can be found in Algorithm~\ref{algo:alg1}.

\begin{algorithm2e*} [!ht]
	\DontPrintSemicolon 
	\KwIn{A set of points $P$ having $A$ attributes, iteration $t$, number of bits to repesent each attribute $k$, number of pairs to be retrieved $s$, number of random coordinates $c$, lower bound of the size of the priority queue $r$}
	\KwOut{Best $s$ pair of points}
	
	\For{each attribute $a_{j} \in A$} {
		
		Initialize an array $S$ of length $A$;
		
		\For{each point $p_{i} \in P$} {
			Let the attribute value of $a_{j}$ in $p_i$ be $v_{i}^{j}$; \\
			Insert the attribute value $v_{i}^{j}$ in $S$;			
		}
	
	    Cluster the values in $S$ into $k$ cluster using any suitable clustering algorithm such as, $k$-means++;
	    
	    \For{each point $p_{i} \in P$} {
	    	Transform attribute value $v_{i}^{j}$ into $k$ binary bits all being turned ``off"; \\
	    	Turn the corresponding bit ``on" according to the cluster label of $v_{i}^{j}$;		
	    }
	}
	
	Initialize a priority queue $Q$ with $r$ keys all being equal to $\infty$; \\
	Initialize a hash table $H$;
	
	\For{$iteration \gets 1$ \textbf{to} $t$} {
		
		\For{each point $p_{i} \in P$} {
			Pick $c$ coordinates randomly from transformed binary space of $p_{i}$; \\
		    Hash the point $p_{i}$ based on these $c$ coordinates in the hash table $H$;
	    }
		
		\For{each pair $m$ of points in each bucket $h_{i} \in H$} {
			
			\If{Hamming distance $d$ (across all the coordinates) between the points in $m$ is $<$ the largest key $\ell$ in $Q$} {
				Delete the pair with the largest key from $Q$; \\
				Insert $m$ into $Q$ with the Hamming distance $d$ as the key;
			}\ElseIf{Hamming distance $d$ (across all the coordinates) between the points in $m$ is $=$ to the largest key $\ell$ in $Q$ and elements in $Q < r$} {
		     	Insert $m$ into $Q$ with the Hamming distance $d$ as the key;
		    }
		}	
	}

    Identify $s$ best pairs (in terms of Euclidean dist. or Pearson's coeff.) in $Q$; \\
	\Return{$s$ best pair of points}; \\
	\vspace{10pt}
	\caption{MSPP: \textbf{M}ining \textbf{S}imilar \textbf{P}airs of \textbf{P}oints}
	\label{algo:alg1}
\end{algorithm2e*}

\subsection{An analysis of our algorithm}
Let $(p',p'')$ be any pair of points. Let $q_1$ be the number of attributes for which both $p'$ and $p''$ have the same encoding. Let $q'$ be the number of coordinates randomly chosen for hashing. Probability  that $p'$ and $p''$ fall into the same bucket is $\frac{{{q_1k}\choose{q'}}}{{{nk}\choose{q'}}}$. Using the fact that ${{a}\choose{b}}\approx\left(\frac{ae}{b}\right)^b$, the above probability is $\approx \left(\frac{q_1}{n}\right)^{q'}$. This in turn implies that the probability that $p'$ and $p''$ do not fall into the same bucket in $z$ successive sampling and hashing steps is $\approx\left[1-\left(\frac{q_1}{n}\right)^{q'}\right]^z$. Using the fact that $\left(1-x\right)^{1/x}\leq 1/e$, for any $1>x>0$, the above probability is $\leq \exp\left[-\left(\frac{q_1}{n}\right)^{q'}z\right]$. This probability will be $\leq n^{-\alpha}$ if $z\geq\alpha\left(\frac{n}{q_1}\right)^{q'}\log n$. Thus we get the following Lemma.

\begin{lemma}
	If a pair $(p',p'')$ of points have the same encoding in $q_1$ of the attributes and if we repeat the sampling and hashing step $\geq\alpha\left(\frac{n}{q_1}\right)^{q'}\log n$ times (choosing $q'$ random columns in every sampling step), then the pair $(p',p'')$ will fall into the same bucket at least once with a probability of $\geq\left(1-n^{-\alpha}\right)$.
\end{lemma}

Then above lemma establishes that if two points are very close to each other in the encoded space, then with a high probability they will form a candidate pair. This in particular means that if the closest pair of points in the original input has a small (not necessarily the smallest) distance in the encoded space, then it will be output as the winner. Now we have to establish this claim. 

Let $\delta$ be the Euclidean distance between the closest pair of points in the original space. There are $n$ attributes. The expected difference between these two points in each of the coordinates is $\frac{\delta}{\sqrt n}$. It follows that in at least $\frac{n}{2}$ attributes, the difference between the two points is $\leq \frac{2\delta}{\sqrt n}$. At the beginning of the algorithm we can randomly pick $\sqrt n$ points, compute the distance between every pair of points in the sample and compute an upper bound $\delta'$ on $\delta$. Using $\delta'$ we can choose the value of $k$ so as to ensure that the closest pair of points will have the same encoding in at least $\frac{n}{2}$ attributes. For example, let $A$ be one of the attributes. Let $d$ be the difference between the maximum and minimum values of the input points with respect to $A$. Refer to $d$ as the deviation in the attribute $A$.  Assume that we cluster the points into $k$ clusters with respect to the attribute $A$. Let $C$ one of the clusters. For any two points in $C$, the distance is clearly, $\leq \frac{d}{k}$. 

Let $d_{\max}$ be the maximum deviation in any of the $n$ attributes. Then we can choose the value of $k$ to be such that: $\frac{d_{\max}}{k}\leq \frac{2\delta'}{\sqrt n}$. In other words, we can choose $k\geq \frac{\sqrt nd_{\max}}{2\delta'}$. Thus we arrive at the following Theorem.

\begin{theorem}
	If we choose the following values for the parameters in the algorithm MSPP, then the closest pair will be output with a probability of $\geq (1-n^{-\alpha})$:
	$k\geq \frac{\sqrt nd_{\max}}{2\delta'}$; $q_1=\frac{n}{2}$; and the number of sampling and hashing steps is $\geq\alpha\left(\frac{n}{q_1}\right)^{q'}\log n$.
\end{theorem}

While we ensure that the closest pair will be a candidate, we also should ensure that not many of the other pairs are candidates. This will ensure that the run time of the algorithm will be under check. One way to ensure this is by letting  the expected number of pairs generated in each sampling step be $O(n)$. Let $q_1$ be the maximum number of attributes in which any pair of points concur (in the encoded space) and $q_2$ be the second largest number of attributes in which any pair of points concur. The probability that the second most correlated pair falls into the same bucket in any specific sampling step  is $\leq \left(\frac{q_2}{n}\right)^{q'}$. If this probability is $\leq\frac{1}{n}$, then the expected number pairs generated in any iteration will be $\leq n$. This happens if $q'=\frac{\log n}{\log(n/q_2)}$.
For this value of $q'$, the number of sampling steps becomes $\geq\alpha n^{\frac{\log(n/q_1)}{\log(n/q_2)}}\log n$.

Given that the expected time we spend in hashing in each iteration and the time for generating the pairs is $O(n)$, it follows that the expected run time of the algorithm (excluding the time taken for data transformation) is $O\left( n^{1+\frac{\log(n/q_1)}{\log(n/q_2)}}\log n\right)$. 

Let $C(n)$ be the time needed to cluster $n$ points. Then, the time taken for data transformation is $m~C(n)$. Thus we get the following Theorem:

\begin{theorem}
	The expected run time of the algorithm MSPP is \[O\left(n^{1+\frac{\log(n/q_1)}{\log(n/q_2)}}\log n+m~C(n)\right).\mbox{\qed}\] 
\end{theorem}
\section{Results and discussions}\label{results}
We have performed rigorous experiments to demonstrate the effectiveness of our proposed algorithm MSPP. These experimental results show that MSPP is indeed efficient, reliable, and scalable which are illustrated next.

\subsection{Datasets}
We have employed both real and synthetic datasets in our experiments. Real datasets were taken from both biomedical and data mining domains. Synthetic datasets were created by randomly generating varying numbers of points and attributes. Next, we provide the details about the datasets.

\subsubsection{Real datasets.} 

To demonstrate the effectiveness of MSPP, we have used $6$ real microarray gene expression datasets. A microarray is a laboratory tool used to detect the expression of thousands of genes simultaneously where each expression of a gene is a real valued number. Each row of a gene expression dataset corresponds to an individual where each column represents the expression of a particular gene across the individuals. Consequently, in our experiment each gene is synonymous with a point and its expressions from different individuals correspond to the distinct attribute values. More details about the datasets can be found in~\cite{zhu}. In addition, another experiment has been carried out to evaluate the performance of MSPP algorithm by employing ``individual household electric power consumption timeseries data" \cite{lich}. 

\subsubsection{Synthetic datasets.}

To perform rigorous simulations, we have generated numerous synthetic datasets by varying the number of points as well as attributes. To mimic the real world scenario, values of a particular attribute are randomly generated using Gaussian distribution having mean $0$ and standard deviation $1$. 



\subsection{Evaluation metrics}

We measure the effectiveness of our proposed algorithm MSPP using $4$ different metrics. These metrics are defined below.

\begin{enumerate}
	
	\item \textbf{A-Rank:} A-Rank means average rank. We have computed the average rank of the top pairs of points detected by algorithm of interest over $5$ runs.
	
	\item \textbf{Accuracy:} Fractions of the pairs of points correctly identified in the top 50 and top 100 pairs of points.
	
	\item \textbf{Speed-up:}  Measures the improvement in execution time of MSPP with respect to other similar algorithms of interest where both the algorithms perform the same task in an identical environment. 
	
	\item \textbf{Time:} Measures elapsed time using total number of CPU clock cycles consumed by each of the algorithms of interest. 
	
\end{enumerate}

\subsection{Outcomes}

\subsubsection{Real datasets.}

We have performed rigorous experimental evaluations to test the scalability, efficiency, and effectiveness of MSPP. As described above our algorithm {\em MSPP} has been run on 6 real gene expression microarray datasets as shown in Table~\ref{tab:tab1}. Both CPU times and accuracy have been used as metrics for our evaluation. Since our algorithm MSPP may not always output the closest pair, we wanted to check the quality of output from our algorithm. To measure this quality, we have used the brute force algorithm to identify the top pairs (the closest, the second closest, the third closest, etc.). We used these outputs to identify the rank of the best output from MSPP. We have also computed the average rank over 5 runs. The results are reported in Table~\ref{tab:tab1}. For all of the cases, MSPP finds the closest pair. Please, note that TSMM algorithms work on time-series data and detect only closest pair of points.

\begin{table}
{\footnotesize
	\caption{Benchmark for MSPP algorithm along with accuracy on real datasets.}\label{tab:tab1}
	\resizebox{\textwidth}{!}{
	\begin{tabular}{cccc|c|c|c|c|c|}
	\cline{1-9}
	\multicolumn{1}{|l}{}                  & \multicolumn{1}{l}{}                        & \multicolumn{1}{l}{}                 & \multicolumn{1}{l|}{} & \multicolumn{2}{c|}{\textbf{CPU time (s)}}   & \multicolumn{3}{c|}{\textbf{Accuracy}}               \\ \hline
	\multicolumn{1}{|l|}{\textbf{Dataset}} & \multicolumn{1}{c|}{\textbf{Name}}          & \multicolumn{1}{c|}{\textbf{Points}} & \textbf{Attributes}   & \multicolumn{1}{l|}{\textbf{Brute-force}} & \textbf{MSPP} & \textbf{A-Rank} & \textbf{Top-50} & \textbf{Top-100} \\ \hline
	\multicolumn{1}{|c|}{D1.1}             & \multicolumn{1}{c|}{Colon Tumor}            & \multicolumn{1}{c|}{2,000}           & 60                    & \textbf{0.49}                             & 0.85 			& 1               & 1.00	          & 1.00           \\ \hline
	\multicolumn{1}{|c|}{D1.2}             & \multicolumn{1}{c|}{Central Nervous System} & \multicolumn{1}{c|}{7,129}           & 60                    & 20.15                                     & \textbf{8.86} & 1               & 1.00              & 1.00           \\ \hline
	\multicolumn{1}{|c|}{D1.3}             & \multicolumn{1}{c|}{Leukemia}               & \multicolumn{1}{c|}{7,129}           & 72                    & 23.70                                     & \textbf{10.20}& 1               & 1.00              & 1.00           \\ \hline
	\multicolumn{1}{|c|}{D1.4}             & \multicolumn{1}{c|}{Breast Cancer}          & \multicolumn{1}{c|}{24,481}          & 97                    & 199.75                                    & \textbf{6.61} & 1               & 1.00              & 1.00           \\ \hline
	\multicolumn{1}{|c|}{D1.5}             & \multicolumn{1}{c|}{Mixed Lineage Leukemia} & \multicolumn{1}{c|}{12,582}          & 72                    & 56.41                                     & \textbf{38.37}& 1               & 1.00              & 1.00           \\ \hline
	\multicolumn{1}{|c|}{D1.6}             & \multicolumn{1}{c|}{Lung Cancer}            & \multicolumn{1}{c|}{12,600}          & 203                   & 138.79                                    & \textbf{96.08}& 1               & 1.00              & 1.00           \\ \hline
	\end{tabular}}%
}
\end{table} 

\begin{table}
	{\footnotesize
		\caption{CPU times consumed by MSPP algorithm with respect to various input sizes and dimensions}\label{tab:tab2}
		\begin{center}
			\begin{tabular}{|c|c|c|c|}
				\hline
				\textbf{Dataset}    	& \textbf{Points}          & \textbf{Attributes} & \textbf{CPU time (m)} \\ \hline
				\multirow{4}{*}{D2.1} 	& \multirow{4}{*}{100,000} & 500                 & 0.60         \\ \cline{3-4} 
				&                          & 1,000               & 1.21       	\\ \cline{3-4} 
				&                          & 1,500               & 1.90         \\ \cline{3-4} 
				&                          & 2,000               & 2.88         \\ \hline
				\multirow{4}{*}{D2.2} 	& \multirow{4}{*}{200,000} & 500                 & 1.10         \\ \cline{3-4} 
				&                          & 1,000               & 2.35         \\ \cline{3-4} 
				&                          & 1,500               & 3.90         \\ \cline{3-4} 
				&                          & 2,000               & 5.15         \\ \hline
				\multirow{4}{*}{D2.3} 	& \multirow{4}{*}{300,000} & 500                 & 1.83         \\ \cline{3-4} 
				&                          & 1,000               & 3.72         \\ \cline{3-4} 
				&                          & 1,500               & 5.98         \\ \cline{3-4} 
				&                          & 2,000               & 7.79         \\ \hline
				\multirow{4}{*}{D2.4} 	& \multirow{4}{*}{400,000} & 500                 & 2.41         \\ \cline{3-4} 
				&                          & 1,000               & 5.27         \\ \cline{3-4} 
				&                          & 1,500               & 7.44         \\ \cline{3-4} 
				&                          & 2,000               & 10.86        \\ \hline
				\multirow{4}{*}{D2.5} 	& \multirow{4}{*}{500,000} & 500                 & 3.41         \\ \cline{3-4} 
				&                          & 1,000               & 6.92         \\ \cline{3-4} 
				&                          & 1,500               & 9.84         \\ \cline{3-4} 
				&                          & 2,000               & 14.62        \\ \hline
			\end{tabular}
		\end{center}
	}
\end{table} 

\begin{table}
	\centering
	\caption{CPU times consumed by MSPP algorithm on large dimensions}\label{tab:tab3}
	\begin{tabular}{|c|c|c|c|}
		\hline
		\textbf{Dataset}    & \textbf{Points}          & \textbf{Attributes} & \textbf{CPU time (m)} \\ \hline
		\multirow{4}{*}{D3.1} & \multirow{4}{*}{500,000} & 500               & 3.19                               \\ \cline{3-4} 
		&                          & 1,000              & 6.85                              \\ \cline{3-4} 
		&                          & 1,500              & 11.11                              \\ \cline{3-4} 
		&                          & 2,000              & 14.10                              \\ \hline
		\multirow{4}{*}{D3.2} & \multirow{4}{*}{1,000,000} & 500               & 9.48                               \\ \cline{3-4} 
		&                          & 1,000              & 17.28                              \\ \cline{3-4} 
		&                          & 1,500              & 25.08                              \\ \cline{3-4} 
		&                          & 2,000              & 34.90                              \\ \hline
		
	\end{tabular}
\end{table} 

\begin{figure}[!ht]
	\subfloat[CPU time consumed by MSPP with respect to varied points and dimensions]{\includegraphics[width = 1.6 in, height = 1.5 in]{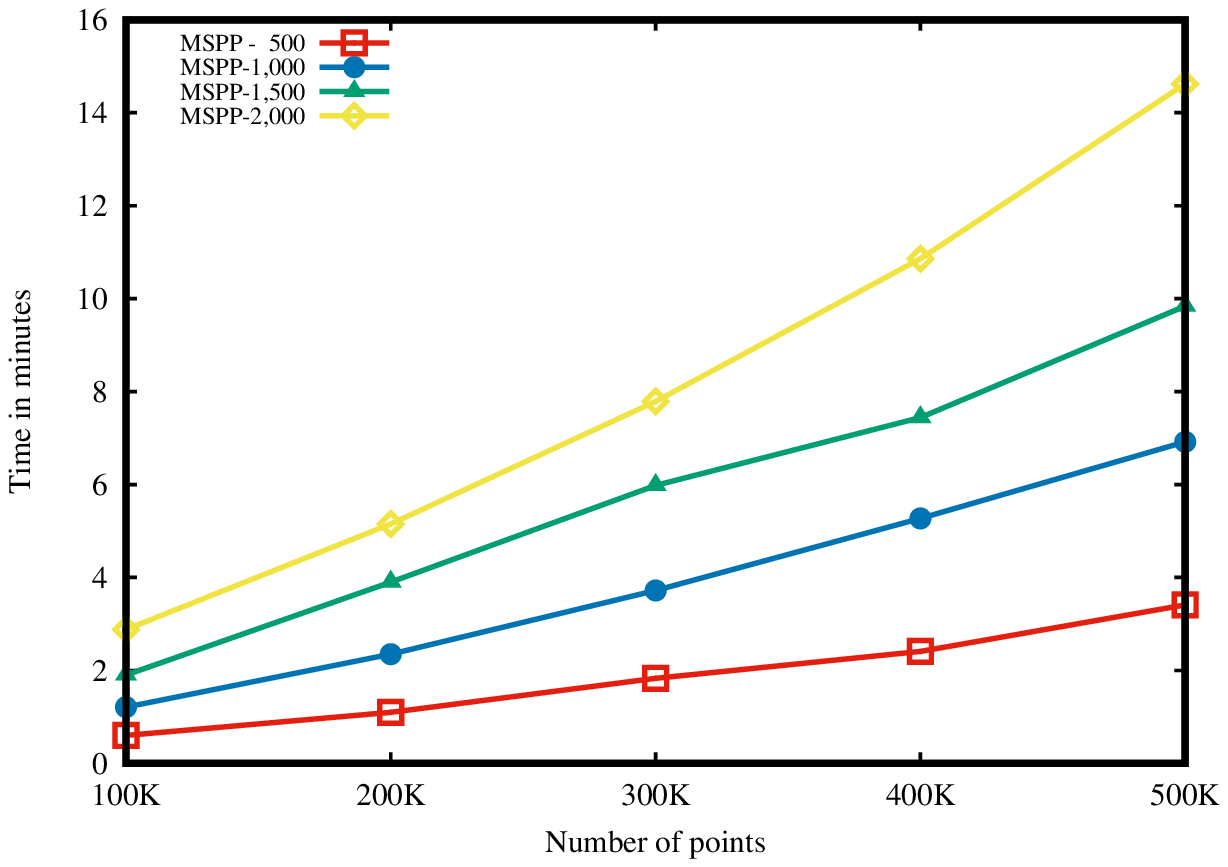}} 
	\subfloat[CPU time consumed by MSPP and JUMP on large datasets]{\includegraphics[width = 1.6 in, height = 1.5 in]{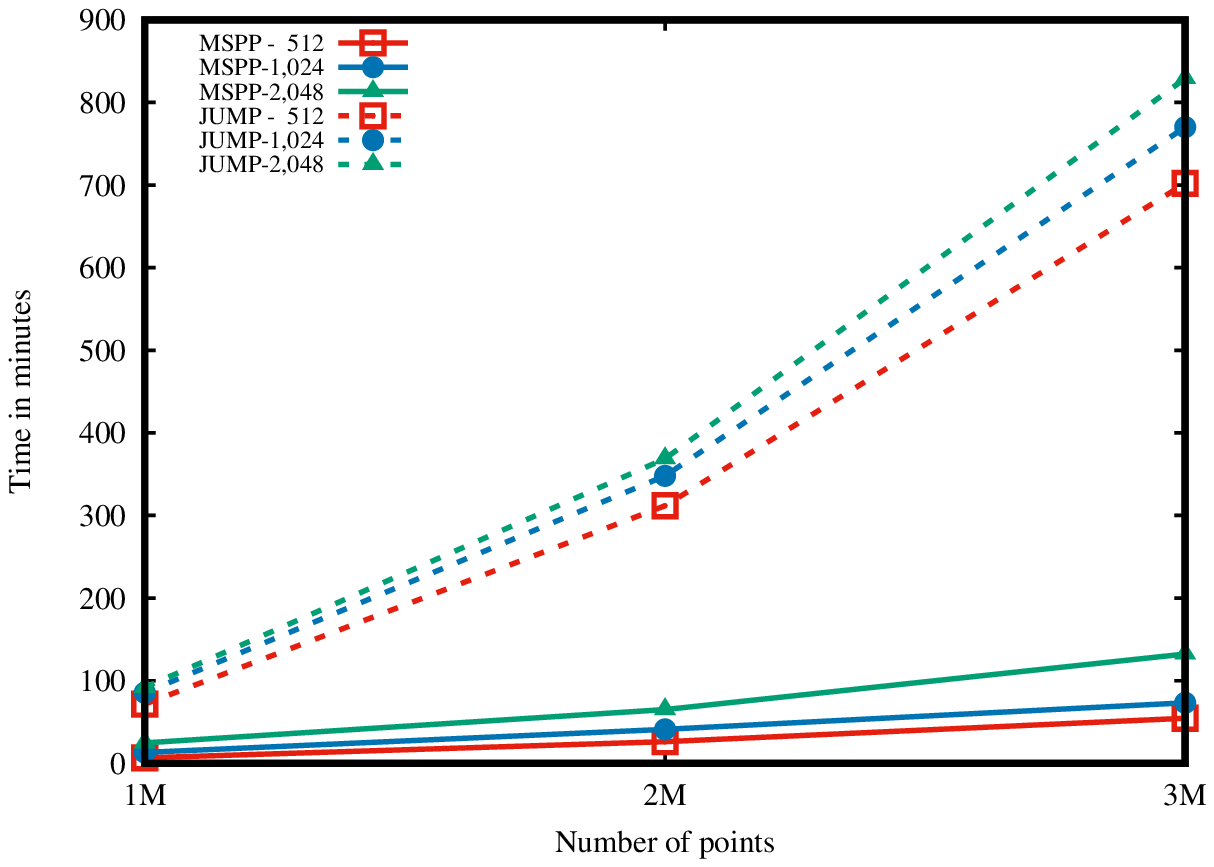}} 
	\subfloat[CPU time consumed by MSPP on large dimensions]{\includegraphics[width = 1.6 in, height = 1.5 in]{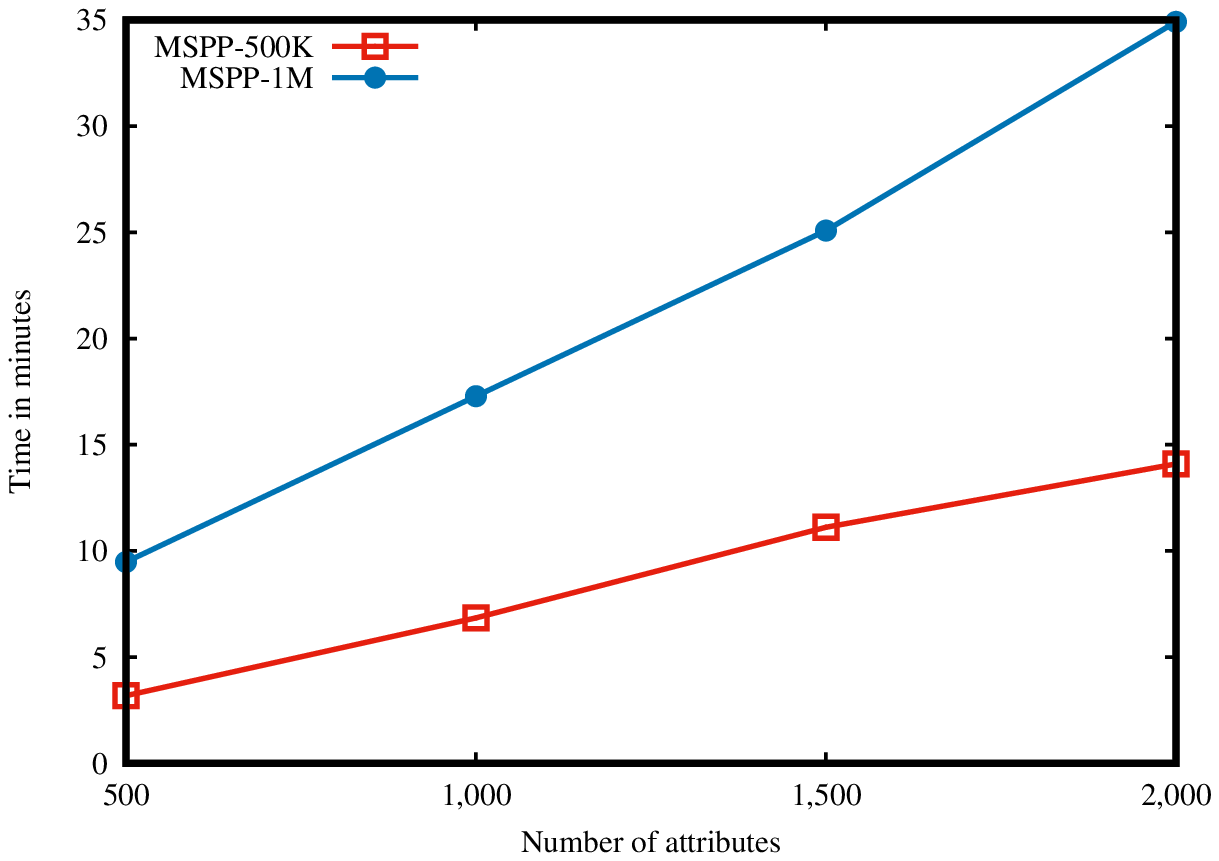}} \\
	\subfloat[CPU time consumed by MSPP and Brute-force]{\includegraphics[width = 1.6 in, height = 1.5 in]{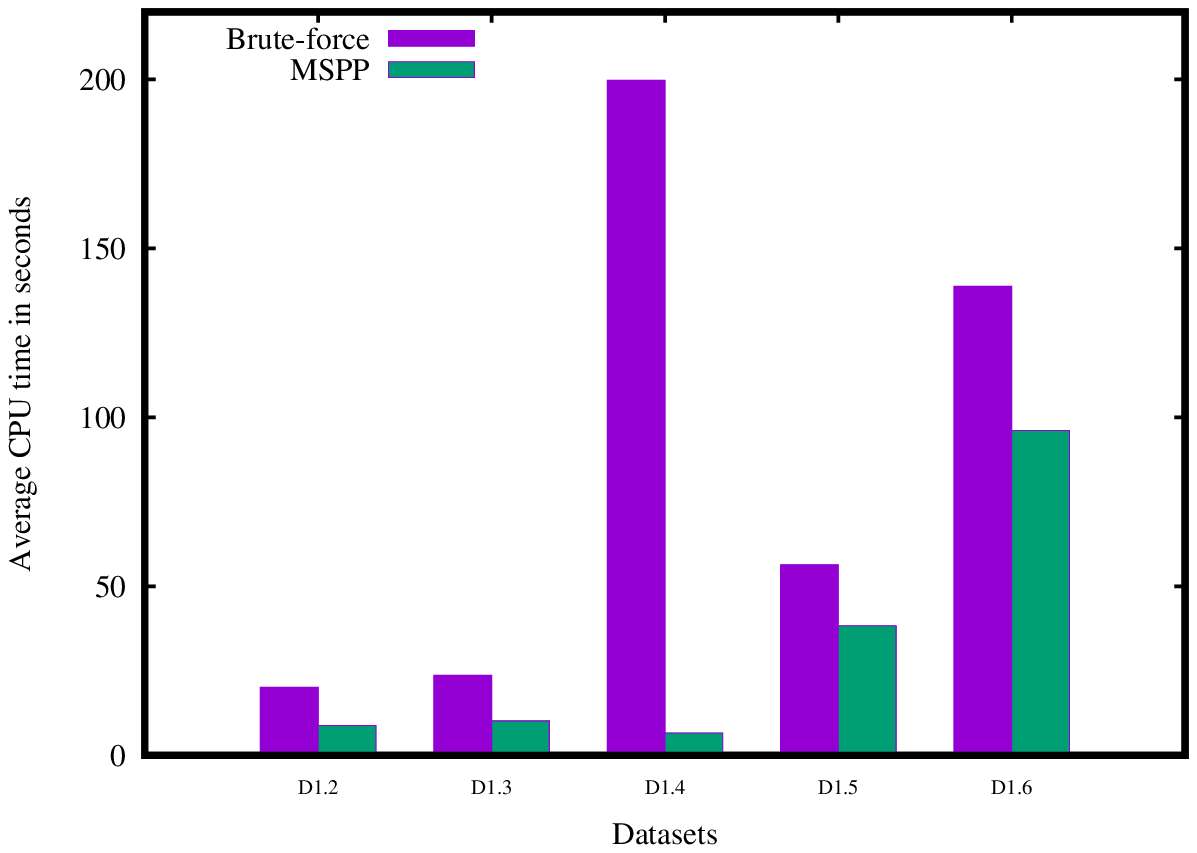}} 
	\subfloat[Speed-up achieved by MSPP over JUMP on large datasets]{\includegraphics[width = 1.6 in, height = 1.5 in]{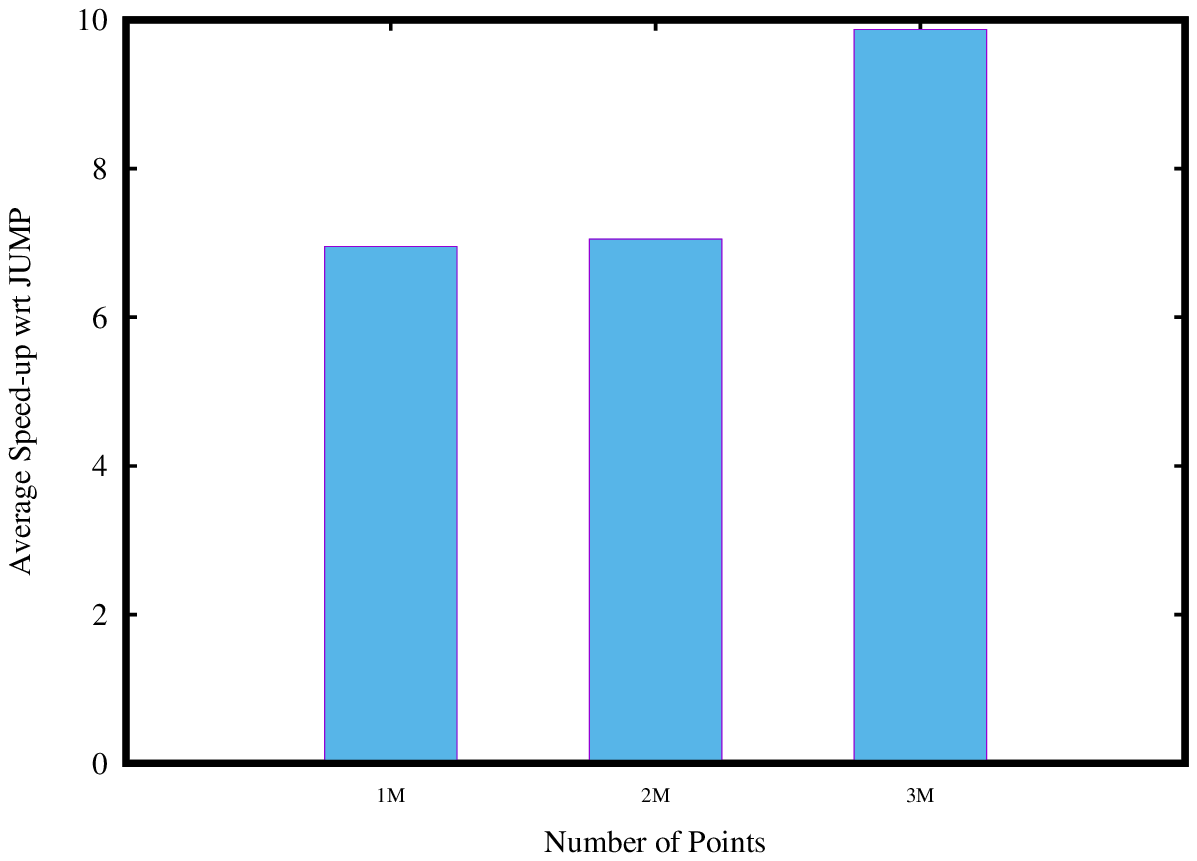}} 
	\subfloat[CPU time consumed by MSPP and Brute-force on randomly picked points from a timeseries]{\includegraphics[width = 1.6 in, height = 1.5 in]{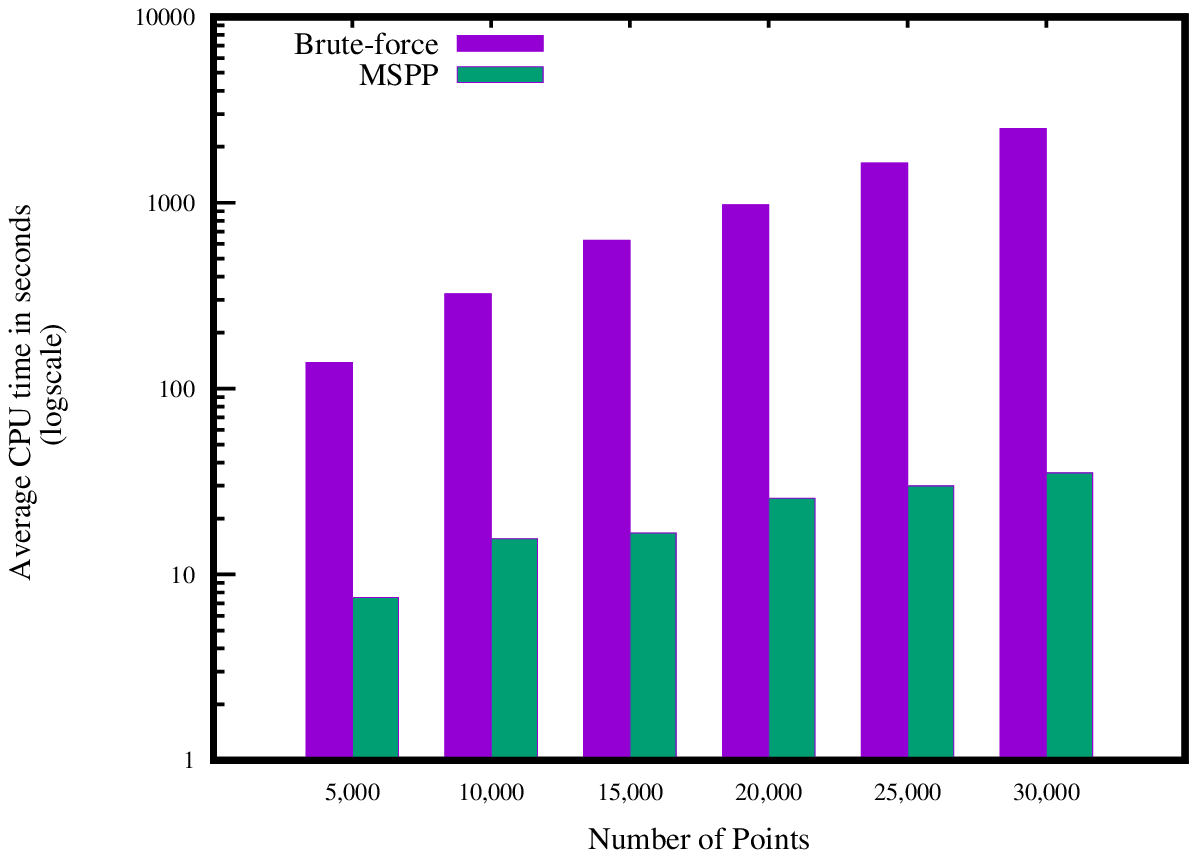}}
	\caption{Performance evaluations of MSPP algorithm.}
	\label{fig:1}
\end{figure}

MSPP not only detects the closest pair of points but also outputs a user defined number of closely similar pairs of points. We have demonstrated it by observing the top 50 and top 100 closest pairs of points from both the brute force and MSPP algorithms. In datasets D1.1 and D1.2 MSPP finds almost all the top 50 and top 100 closest pairs of points. It identifies all the top 50 and top 100 closest pairs in D1.4-D1.6 datasets. In dataset D1.3 it recognizes nearly half of the top 50 and top 100 closest pairs. The improvement of performance in terms of execution time of MSPP becomes more significant for larger datasets such as in D1.4. Please, see Table~\ref{tab:tab1} and Figure~\ref{fig:1}(d) for runtime comparisons. In each experiment 10 bits were used to encode each attribute value. MSPP picked 20 coordinates randomly in each stage of sampling and the number of stages was 5.

Now, consider the timeseries dataset. The experiments were done in a similar fashion as stated above. On each run a number of points were randomly picked from the timeseries and then, the MSPP and brute-force algorithms were executed on these points. Each point has a length of $1,000$ attributes. Results of this experiment are summarized in Table~\ref{tab:tab9}. For a visual comparison please see Figure~\ref{fig:1}(f). Please note that a log scale has been used for the y-axis. 

\begin{table}[!ht]
	{\footnotesize
		\caption{CPU times consumed by MSPP and JUMP on large datasets along with speed-up ratios.}\label{tab:tab4}
		\begin{center}
			\begin{tabular}{|c|c|c|c|c|c|c|}
				\hline
				\multirow{2}{*}{\textbf{Dataset}} & \multirow{2}{*}{\textbf{Points}} & \multirow{2}{*}{\textbf{Attributes}} & \multicolumn{2}{|c|}{\textbf{CPU time (m)}} & \multicolumn{2}{|c|}{\textbf{Speed-up}} \\ \cline{4-7}
				& 								 & 									   & \textbf{JUMP}    & \textbf{MSPP}         & \textbf{Ratio}     & \textbf{Average} \\ \hline
				\multirow{3}{*}{D4.1} & \multirow{3}{*}{1,000,000} & 512   & 71.67  & \textbf{6.74}   & 10.63 & \multirow{3}{*}{6.95} \\ \cline{3-6}
				&                            & 1,024 & 85.59  & \textbf{13.15}  & 6.51  &                       \\ \cline{3-6}
				&                            & 2,048 & 91.78  & \textbf{24.82}  & 3.70  &                       \\ \cline{3-6}
				\hline
				\multirow{3}{*}{D4.2} & \multirow{3}{*}{2,000,000} & 512   & 311.86 & \textbf{26.29}  & 11.86 & \multirow{3}{*}{7.05} \\ \cline{3-6}
				&                            & 1,024 & 348.04 & \textbf{41.29}  & 8.43  &                       \\ \cline{3-6}
				&                            & 2,048 & 368.86 & \textbf{65.07}  & 5.67  &                       \\ \cline{3-6}
				\hline
				\multirow{3}{*}{D4.3} & \multirow{3}{*}{3,000,000} & 512   & 701.96 & \textbf{54.67}  & 12.84 & \multirow{3}{*}{9.87} \\ \cline{3-6}
				&                            & 1,024 & 770.06 & \textbf{73.27}  & 10.51 &                       \\ \cline{3-6}
				&                            & 2,048 & 829.01 & \textbf{132.45} & 6.26  &                       \\ \cline{3-6}
				\hline
			\end{tabular}
		\end{center}
	}
\end{table} 

\begin{table}[!ht]
	{\footnotesize
		\caption{Benchmark for MSPP on randomly picked points with $1,000$ attributes from the entire space of points from the timeseries data.  }\label{tab:tab9}
		\begin{center}	
			\begin{tabular}{|c|c|c|c|c|c|}
				\cline{1-6}
				\multicolumn{1}{|c|}{}                &   \multicolumn{2}{c|}{\textbf{Average CPU time (s)}}   & \multicolumn{3}{c|}{\textbf{Accuracy}}               \\ \hline
				\multicolumn{1}{|c|}{\textbf{Points}} &  \multicolumn{1}{c|}{\textbf{Brute-force}} & \textbf{MSPP} & \textbf{A-Rank} & \textbf{Top-50} & \textbf{Top-100} \\ \hline
				
				\multicolumn{1}{|c|}{5,000} & \multicolumn{1}{c|} {137.87} & \textbf{7.52}  & 1 & 1.00 & 1.00 \\ \hline
				\multicolumn{1}{|c|}{10,000} & \multicolumn{1}{c|}{323.66} & \textbf{15.59} & 1 & 1.00 & 1.00 \\ \hline
				\multicolumn{1}{|c|}{15,000} & \multicolumn{1}{c|}{627.74} & \textbf{16.74} & 1 & 1.00 & 1.00 \\ \hline
				\multicolumn{1}{|c|}{20,000} & \multicolumn{1}{c|}{973.73} & \textbf{25.73} & 1 & 1.00 & 0.99 \\ \hline
				\multicolumn{1}{|c|}{25,000} & \multicolumn{1}{c|}{1,636.02}& \textbf{29.93} & 1 & 0.94 & 0.93 \\ \hline
				\multicolumn{1}{|c|}{30,000} & \multicolumn{1}{c|}{2,510.25}& \textbf{35.19} & 1 & 0.85 & 0.84 \\ \hline
				
			\end{tabular}
		\end{center}
	}
\end{table} 

\subsubsection{Synthetic datasets.}

A set of experiments have been carried out using randomly generated data (as pointed out earlier) to evaluate the performance of MSPP. To study the effect of input sizes and attributes on the performance, a total of $20$ datasets have been generated with varying numbers of points and attributes as shown in Table~\ref{tab:tab2}. The execution times to identify the closely similar pairs of points from these datasets are reported in Table~\ref{tab:tab2}. Please, see Figure~\ref{fig:1}(a) for visual details. It is evident that the run time almost linearly increases with the number of points. To further study the effect of large dimensions on our algorithm, another set of datasets have been generated by varying the number of dimensions. In these datasets the input size is fixed at 100,000 points while the dimension (number of attributes) is increased from 5,000 to 20,000 in steps of 5,000 increment (Please, see Table~\ref{tab:tab3}). The execution times are illustrated in Figure~\ref{fig:1}(c). This set of experiments reveal the linear relationship between the execution time of MSPP and the number of dimensions.

Time series motif mining (TSMM) is a crucial problem that can be thought of as a special case of the CPP in a large dimensional space as described earlier. Since, there are very efficient algorithms (e.g. \cite{ref_jump,ref_ll}) in the literature for solving the TSMM problem, we have further conducted experiments to investigate their performances on simulated points generated by employing Gaussian distribution as stated earlier. This time the number of attributes has been fixed at $1,000$ and four large datasets have been generated (please, see Table~\ref{tab:tab4}). The runtimes and corresponding speed-up ratios achieved by MSPP over JUMP are recorded in Table~\ref{tab:tab4}. The execution times for both algorithms have been plotted in Figure~\ref{fig:1}(b) while Figure~\ref{fig:1}(e) shows the speed-up ratios. These comparisons reveal that MSPP is faster than JUMP. For example, MSPP is almost $10 \times$ faster than JUMP on the D4.3 dataset. Please note that we also tried to include the execution times of the LL algorithm in this comparative study. It has been observed that the LL algorithm requires more than 70 hours to run on $1,000,000$ points with $1,000$ attributes and so, we are not reporting the runtimes of LL. In each experiment for the synthetic datasets 2 bits were used to encode each attribute value. MSPP picked 20 coordinates randomly in each stage of sampling and the number of stages was 5. It is to be noted that any time series motif mining algorithm will only identify the closest pair from the given set of points. On the contrary, MSPP identifies a user defined number of similar pairs of points. In the above experiments, MSPP outputs 500,000 similar pairs of points containing the closest pair with a high probability. In this respect, a direct comparison may not be appropriate between MSPP and any other time series motif miner. 

\section{Some notable applications of MSPP}
\label{apps}

MSPP can be used in diverse set of practical applications such as but not limited to time series motif mining, clustering, gene co-expression network, feature reduction in high dimensional space, 2-locus problem in genome-wide association study (GWAS), outliers detection, and so on. Next we briefly discuss two of the notable applications of MSPP because of page constraints.  

\subsection{Scalable clustering}

Clustering is the task of bundling a set of objects into groups of similar objects and it is an example of unsupervised learning. The ever-increasing size of datasets and poor scalability of clustering algorithms are some of the bottlenecks for unsupervised learning. We can employ MSPP to ensure scalability as well as speed-up any centroid based clustering algorithms such as, $k$-means and its variations. As MSPP detect similar pair of points, we can use those points as a surrogate of the entire dataset. Suppose, we know a subset of points with cluster labels \emph{a priory}. If we have enough representative points (i.e., points having very close pair-wise distances) from a cluster, we can compute approximate centroid mimicking the original one. Then we can label rest of the points with unknown class labels based on those centroids. In this context MSPP gives us a subset of very closely similar pair of points. We cluster those points using $k$-means algorithm and extract the centroids. At the end we assign cluster labels to the rest of the points based on those centroids. I.e., for each point we compute Euclidean distances between this point to every centroids and assign it to that centroid having the least distance.   

\begin{table}[!ht]
	{\footnotesize
		\caption{Clustering performance by employing MSPP}\label{tab:tab5}
		\begin{center}
			\begin{tabular}{|c|c|c|c|c|} \hline
				\textbf{Dataset}  & \textbf{Points} & \textbf{Clusters}  & \textbf{Sample points}  &  \textbf{Accuracy} \\ \hline
				D5.1 & 5,000 & 5 & 1,808 & 1.0\\ \hline
				D5.2 & 10,000 & 5 & 3,325 & 1.0\\ \hline
				D5.3 & 15,000 & 10 & 5,037 & 1.0\\ \hline
				D5.4 & 20,000 & 10 & 6,941 & 1.0\\ \hline
			\end{tabular}
		\end{center}
	}	
\end{table}

To demonstrate the effectiveness and scalability of this method we randomly populate datasets by varying number of points and clusters using Gaussian distribution. At the beginning MSPP runs on those data points and detects a subset of similar pair of points in Euclidean space. Let the particular dataset contain $k$ clusters. We cluster those points into $k$ clusters and assign the rest based on the centroids as described above. Experimental evaluations show that we correctly assign cluster labels to all of the data points in various datasets. It suggests the potential of MSPP algorithm to make a clustering method scalable. Please, see Table \ref{tab:tab5} for more details. Here, accuracy is the fraction of data points correctly labeled. 

\subsection{Dimension reduction}

There are a number of ways in which any two human genomes can differ. Variations are largely due to the single nucleotide polymorphisms (SNPs, in short) as well as deletions, insertions, and copy number variations. Any of these variations may result in alterations in an individual's traits, or phenotypes. Investigations that try to understand human variability using SNPs fall under genome-wide association study (GWAS). In a typical GWAS 85-100 million of SNPs are sequenced from not more than several thousands individuals. In addition to this, SNPS are often highly correlated. In general correlated features (such as, SNPs) do not improve the machine learning model of interest. It depends on the various factors of the problem we are solving such as the number of variables and the degree of correlation among the features. It may affect specific models in different ways and to varying extents. There are basically three main reasons to remove correlated features from the set of given features: (1) making the learning algorithm faster; (2) decreasing harmful bias; and (3) making the model simpler and interpretable.

By carefully discarding highly correlated SNPs we can ultimately reduce the dimension of the problem. In this context, we can employ MSPP algorithm to construct a graph of highly correlated SNPs where each node in the graph will act as a SNP. There will be an edge between two nodes if and only if they are highly correlated. By carefully analyzing the graph we can select a subset of SNPs and discard the rest. For an example, if we find a clique in the graph we can retain one node and discard the rest from that graph.

\section{Conclusions}
\label{conclusion}

In this article we have proposed an efficient, reliable, and scalable algorithm called MSPP to detect a set of highly similar pairs of points in both of Euclidean and Pearson's metric spaces. It is an out-of-core algorithm and thus, it can work on a large high dimensional dataset. MSPP consumes less amount of physical memory. There are two basic steps in the algorithm and in each step, we do one pass through the entire data. In each pass the algorithm incrementally retrieves information embedded in the dataset by discretizing and encoding attributes. After the final pass, it outputs a set of similar pairs of points. Experimental evaluations show that MSPP is indeed reliable, scalable, and efficient in terms of both accuracy and execution time. MSPP can be used in a diverse set of practical applications such as, but not limited to, time series motif mining, clustering, gene co-expression network, feature reduction in high dimensional space, and 2-locus problem in genome-wide association study.

\bibliographystyle{splncs04}

\bibliography{ms}

\end{document}